\documentclass[]{svjour3}

\usepackage{graphicx}
\usepackage[]{natbib}
\usepackage{amsmath}
\usepackage{amssymb}

\def\E{\mathcal{E}} 
\def\R{\mathcal{R}} 
\def\Q{\mathcal{Q}}

\newcommand{\PP}{\mathbb{P}}

\begin{document}

\title{Autocatalytic Sets and Biological Specificity}
\author{Wim Hordijk \and Peter R. Wills \and Mike Steel}
\institute{Wim Hordijk \at
           SmartAnalytiX.com \\
           Lausanne, Switzerland\\
           \email{wim@WorldWideWanderings.net}
         \and
           Peter R. Wills \at
           Universit\"{a}t T\"{u}bingen, Intergrative Transcriptomics\\
           T\"{u}bingen, Germany\\
           and\\
           University of Auckland, Dept. of Physics\\
           Auckland, New Zealand\\
           \email{p.wills@auckland.ac.nz}
         \and
           Mike Steel \at
           University of Canterbury, Allan Wilson Centre for Molecular
           Ecology and Evolution\\
           Christchurch, New Zealand\\
           \email{mike.steel@canterbury.ac.nz}}
\date{Received: date / Accepted: date}
\maketitle

\begin{abstract}
A universal feature of the biochemistry of any living system is that all the molecules and catalysts that are required for reactions of the system can be built up from an available food source by repeated application of reactions from within that system. RAF (reflexively autocatalytic and food-generated) theory provides a formal way to study such processes. Beginning with Kauffman's notion of ``collectively autocatalytic sets'', this theory has been further developed over the last decade with the discovery of efficient algorithms and new mathematical analysis. In this paper, we study how the behaviour of a simple binary polymer model can be extended to models where the pattern of catalysis more precisely reflects the ligation and cleavage reactions involved. We find that certain properties of these models are similar to, and can be accurately predicted from, the simple binary polymer model; however, other properties lead to slightly different estimates. We also establish a number of new results concerning the structure of RAFs in these systems.
\end{abstract}

\section{Introduction}

In its broadest sense, the term ``autocatalysis'' refers to a process whereby some entity facilitates the chemical construction of another instance of itself.  Because this is also a molecular-level description of biological reproduction, the study of simply specified autocatalytic systems has proven to be a fruitful field for gaining insights into possible origins of life. While \citet{Eigen:71} described the selective preservation of information in systems involving macromolecular sequences undergoing competitive reproduction, \citet{Kauffman:71,Kauffman:86} drew attention to the coincidence of cooperative catalytic functionalities that could potentially create a self-sustaining system of polymers, irrespective of whether they resembled information-carrying genes, the hallmark of quasi-species and hypercycles \citep{Eigen:71,Eigen:79}. The disparity between these approaches reflects differing views of how we account for the complex chemistry of biological systems. What features of molecular biological and biochemical processes characterise their integration at the origin of life? For \citet{Eigen:71}, the answer lies in the capacity of replicating polymers like RNA and DNA to evolve as a result of Darwinian selection, whereas \citet{Kauffman:71,Kauffman:86} urges us to look at the possibility of self-amplifying networks generating themselves as a result of nothing more than natural coincidences of connectedness. It is not our intention to adjudicate the dispute implicit in these divergent points of view. Rather, we now wish to investigate the extent to which the theory of autocatalytic sets and its recent extensions \citep{Kauffman:71,Kauffman:86,Kauffman:93,Steel:00,Hordijk:04,Mossel:05,Hordijk:11,Hordijk:12,Hordijk:12a,Hordijk:12b,Hordijk:13} can be applied to some of the core ideas of molecular biology and thereby contribute to the incremental refinement of the problem which is usually posed as the unanswerable question ``What is Life?"

A significant objection raised against the idea of simple autocatalytic sets forming the nucleus of the original molecular processes that led to biology concerns the question of ``biological specificity''. This was the term that the first molecular biologists (e.g., \citet{Crick:58,Crick:70}) used to articulate the profound impression that variation in the structure of a single molecule (DNA) was responsible for the orderly variation in the corresponding particularities of organisms. The discovery and elucidation of the direct transfer of information from sequences of nucleotide triplets in DNA (or RNA) to sequences of amino acids in proteins -- the genetic code -- followed the prediction of \citet{Schroedinger:44} that an organism, the biological phenotype, is constructed by use of a genotypic ``codescript''. Schr{\"o}dinger envisaged the codescript as information stored in an ``aperiodic crystal", the main structural features of which were found to be met by the one-dimensional sequence of heteropolymeric DNA \citep{Watson:53}.

However, it was soon realised that the mapping from polymer sequence information (especially genes) to the phenotypic properties of organisms was much more convoluted than the coded transfer of information from DNA to RNA to protein. The first step following execution of the genetic code, protein folding, results in the original sequence information becoming scrambled: the algorithm that maps it onto the catalytic functions of proteins is exceedingly complex.   Nevertheless, though scrambled, this mapping is still orderly, rather than random; and if it were completely random there would be no path whereby success in the struggle for survival could be reliably reverse-coded in polymer sequences through a gradual process of adaptation through natural selection. These considerations lead to the conclusion that if autocatalytic sets of polymers are to provide a plausible explanation for the origin of life, then their probable emergence must be demonstrated for systems of polymers like RNA and proteins. In the case of these polymers the variation of catalytic function with sequence displays  an orderliness, albeit scrambled, that satisfies the fundamental principle of molecular chemistry: similar structures tend to have similar properties -- molecular structure-function mappings are far from random. 

The original description of probabilistic autocatalytic sets of polymers \citep{Kauffman:71,Kauffman:86} was susceptible to criticism along these lines: chemical reactions and catalytic functions were enumerated without any regard for the structural (and therefore the likely chemical) similarities that polymer sequences share. Considerations of the role of  complementary sequence matching as a model of substrate recognition in catalytic processes have gone some way to addressing this criticism \citep{Kauffman:93,Hordijk:11,Hordijk:12}, but there is another aspect of specific molecular recognition processes that is not very realistically represented when polymers (of all lengths up to a certain size) are randomly assigned as catalysts of ligation or cleavage reactions. With random assignments it is possible for a molecule comprising only a few atoms to be required to recognise  the exact sequences of two much bigger molecules that it has the task of ligating, or creating through cleavage. In the absence of any other constraints, it is hard to imagine how a polymer sequence of length $k$ could specifically recognise portions of other polymers comprised of many more than $k$ monomers in total. In the most extreme case, how could a monomer or dimer be a ligation catalyst for the ligation of two particular polymers of length significantly greater than 2, but not others with the same end-sequences at the site of ligation or cleavage? This could only occur in a real chemical system if there were other factors, in addition to the direct sequence recognition capabilities of the catalyst, that made those two particular polymers and not others prone to such catalytic ligation or cleavage. Special chemical constraints of this sort cannot be accommodated in a random mapping from polymer sequence to catalytic function.

In this work, we take steps to address this problem. In the first place, we demand that any polymer that acts as a catalyst must contain a structure that is realistically capable of recognising the molecular features on which it acts.  The minimum recognition structure for a catalyst is taken to be an oligomeric sequence complementary (in the two-letter alphabet) to the ligation/cleavage sequence that it acts on. We also consider cases in which the recognition structure may be required to contain more bits of information than the complementary sequence alone, while remaining contiguous with it. This addresses two aspects of chemical realism: (i) that the properties of a particular local structure, as long as it is intact, will not usually be unduly affected by remote structural features; and (ii) that an orderly variation in function correlates with an orderly variation in structure. The second demand we make is that the functional complexity of structures should be commensurate with their structural complexity. We achieve this in the simplest possible way, by making the restriction that only molecules of maximum length can act as catalysts and the maximum sum of features they can recognise is of the same size. This restriction is rather crude but it achieves the desired result without adding elaborate details of indeterminate effect on our elementary model. We consider the effect of these requirements individually and then together. We reach the conclusion that autocatalytic systems that do not involve information storage and coded transfer do not have zero capacity for the maintenance of biochemical specificity, a conclusion at apparent variance with the Sequence Hypothesis of \citet{Crick:58}.

\section{Chemical reaction systems and autocatalytic sets}

We briefly review the relevant definitions and main results of autocatalytic set theory. First, a {\it chemical reaction system} (CRS) is defined as a tuple $Q=\{X,\R,C\}$ consisting of a set of molecule types $X$, a set of chemical reactions $\R$ and a catalysis set $C$ that indicates which molecule types catalyse which reactions. We also include the notion of a food set $F \subset X$, which is a subset of molecule types that are assumed to be freely available from the environment. An {\it autocatalytic set} (or reflexively autocatalytic and food-generated (RAF) set) is now defined as a subset $\R' \subseteq \R$ of reactions and associated molecule types which are:
\begin{enumerate}
\item {\it Reflexively autocatalytic} (RA): each reaction $r \in \R'$ is catalysed by at least one molecule type involved in $\R'$, and
\item {\it Food-generated} (F): all reactants in $\R'$ can be created from the food set $F$ by using a series of reactions only from $\R'$ itself.
\end{enumerate}

\begin{figure}[htb] 
\centering
\includegraphics[scale=0.6]{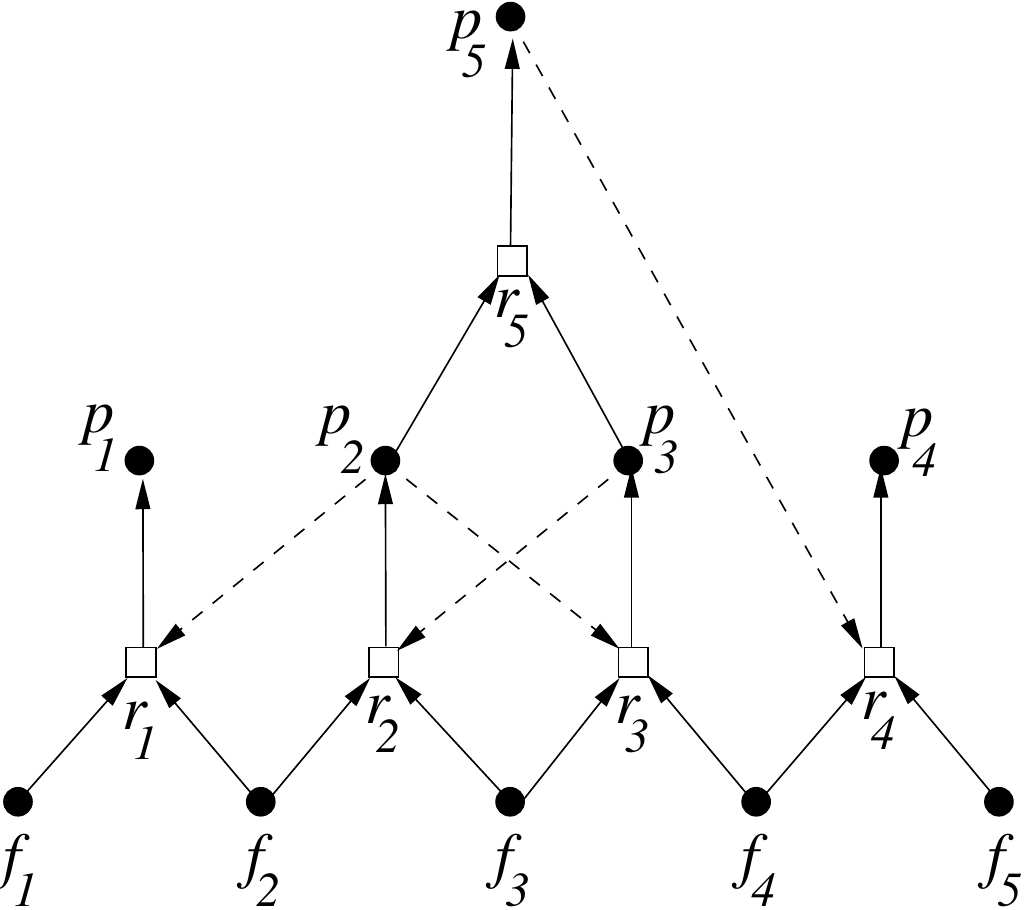}
\caption{A CRS that contains two RAF sets, the maxRAF $\{r_1, r_2, r_3\}$ and the irrRAF $\{r_2, r_3\}$. Here $F= \{f_1, \ldots, f_5\}$, $X=F \cup \{p_1, \ldots, p_5\}$, $\R = \{r_1,\ldots,r_5\}$, and catalysis is indicated by dashed arrows.}
\label{fig:CRS}
\end{figure}

A more formal definition of RAF sets is provided in \citet{Hordijk:04,Hordijk:11}, including an efficient algorithm for finding RAF sets in general chemical reaction systems. It was shown that RAF sets are highly likely to exist in a simple model of chemical reactions systems known as the binary polymer model \citep{Hordijk:04,Mossel:05}, and that this result also holds when more realistic assumptions are included in the model \citep{Hordijk:11,Hordijk:12}. An example of a simple CRS that contains RAF sets of size two and three is shown in Fig. \ref{fig:CRS}.

The RAF sets that are found by the RAF algorithm are called {\it maximal} RAF sets (maxRAFs). However, it turns out that a maxRAF can often be decomposed into several smaller subsets which themselves are RAF sets (subRAFs) \citep{Hordijk:12a}. If such a subRAF cannot be reduced any further without losing the RAF property, it is referred to as an {\it irreducible} RAF (irrRAF). The existence of multiple autocatalytic subsets can actually give rise to an evolutionary process \citep{Vasas:12}, and the emergence of larger and larger autocatalytic sets over time \citep{Hordijk:12a,Hordijk:12b}. Recently, the formal RAF framework was also applied to an experimental chemical system of catalytic RNA molecules in which autocatalytic sets emerged spontaneously \citep{Vaidya:12}. The formal model is capable of reproducing the main experimental results and also provided additional insights and predictions about the system's behaviour \citep{Hordijk:13}.

\section{Models of chemical reaction systems}

Here, we apply the RAF framework to two related models of chemical reaction systems, both of which are variants and extensions of the  binary polymer model used previously.  First, we briefly review the basic model, and then describe the two variants.

\subsection{The binary polymer model}

The binary polymer model was originally introduced by Kauffman in the context of studying autocatalytic sets \citep{Kauffman:86,Kauffman:93}. Polymers are represented by strings of 0s and 1s, and the possible reactions are cleavage and ligation. Catalysis is assigned at random.

\subsubsection{The molecule set}
The molecule set $X$ consists of all bit strings up to (and including) a maximum length $n$:
\[ X = \{0,1\}^{\leq n}.  \]
Therefore, there are $|X| = 2^{n+1}-2$ molecule types.

\subsubsection{The food set}
The food set $F$ consists of all bit strings up to (and including) a certain length $t$:
\[ F = \{0,1\}^{\leq t}. \]
Usually, $t << n$ (e.g., $t=2$ or $t=3$ is used).

\subsubsection{The reaction set}
The reaction set $\R$ consists of all possible ligations (i.e., ways of ``gluing'' two bit strings together without violating the maximum length constraint) and cleavages (breaking a bit string into two parts). An example of a ligation reaction is $000+1111\rightarrow0001111$; one for a cleavage reaction is $010110101\rightarrow0101+10101$.

There are $|\R| = (n-2)2^{n+1}+4$ possible ligation/cleavage reaction pairs, which can also be considered as one bi-directional reaction (although in terms of finding RAF sets, this does not make a difference).

\subsubsection{The catalysis set}
The catalysis set $C$ is made up of combinations of molecules (bit strings) and reactions:
\[ C = \{(x, r) | x \in X, r \in \R \}, \]
In the model, these catalysis events are assigned independently and with equal probability $p(n)$ across all possible $(x, r)$ pairs (there are $|X||R|$ such pairs, where the reactions $r$ are considered to be bi-directional).

\subsubsection{RAF sets}
The binary polymer model was introduced to show that autocatalytic sets are highly likely to exist for a large enough diversity of molecule types, i.e.,  a large enough value of $n$ \citep{Kauffman:86,Kauffman:93}. These arguments and results were refined later on, showing that RAF sets have a high probability of existence even for very moderate levels of catalysis -- between one and two reactions catalysed per molecule, on average, for values of $n$ at least up to 50 \citep{Hordijk:04}. Furthermore, despite the number of reactions growing exponentially with increasing $n$, a growth rate in the level of catalysis that is linear (with increasing n) is sufficient (and also necessary) to maintain a high probability of RAF existence \citep{Hordijk:04,Mossel:05}.

\subsection{An extended binary polymer model}

Consider the following extended version of the binary polymer model
in which the catalysis events $\E(x,r,n)$ (i.e., $x$ catalyses $r$, for a maximum molecule length $n$) are still independent across $x$ and $r$, but where $\Pr[\E(x,r,n)]$ can also depend on (some property of) $x$ and $r$, instead of only on $n$. Allowing arbitrary dependence, however, is problematic. For example, suppose that one molecule $m$ catalyses all reactions, or suppose that all reactions except the ones required in the last step to form $m$ and all other molecules catalyse no reactions. Then the probability of an RAF can be arbitrarily close to 1 or 0, respectively. To obtain a balance between realism and tractability, we consider the following extended model in which $0 \leq p(n) \leq 1$ and $0 \leq m(x,r,n) \leq 1$ for all $x,r,n$:
\begin{equation}
\Pr[\E(x,r,n)] = p(n) \cdot m(x,r,n), 
\label{probex}
\end{equation}
where $m(x,r,n)$ is the probability that $x$ and $r$ conform to a given set of constraints (possibly involving $n$), and $p(n)$ is the probability that $x$ catalyses $r$ {\it given} that they conform to those constraints.

Here, we consider four versions of this extended model:
\begin{itemize}
\item {\tt RAND}: The original (purely random) binary polymer model:
  \[ m(x,r,n) = 1 \]
\item {\tt TMPL}: A template-based catalysis model, where $x$ is considered a candidate catalyst for $r$ only if, somewhere along its sequence, it matches the reaction template of $r$ (or, equivalently, the complement of the reaction template). This reaction template could, for example, consist of the four bits (two on either side) around the cleavage/ligation site. We use the notation $x \sim r$ to indicate such a template match between $x$ and $r$. We thus have:
  \[ m(x,r,n) = \left\{ \begin{array}{ll} 1, & x \sim r; \\
     0, & \mbox{otherwise.} \end{array} \right. \]
\item {\tt MLEN}: Only molecules of maximum length $n$ are considered as candidate catalysts:
  \[ m(x,r,n) = \left\{ \begin{array}{ll} 1, & \mbox{if } |x|=n; \\
     0, & \mbox{otherwise.} \end{array} \right. \]
\item {\tt BOTH}: A combination of the template-based and maximum-length constraints:
  \[ m(x,r,n) = \left\{ \begin{array}{ll} 1, & \mbox{if } x \sim r
     \mbox{ and } |x|=n; \\ 0, & \mbox{otherwise.} \end{array} \right. \]
\end{itemize}

Note that the {\tt RAND} version of the model (i.e., the original model) was already described and investigated in detail in \citet{Kauffman:86,Kauffman:93,Hordijk:04,Mossel:05} and the {\tt TMPL} version (with a four-bit template) in \citet{Hordijk:11,Hordijk:12}. However, we have included these versions here for completeness and comparison (and as specific instances of the more general extended model), while the main interest is in the {\tt MLEN} and {\tt BOTH} versions of the model.

\subsection{The Wills--Henderson Model}

The Wills--Henderson (W-H) model, originally introduced in \citet{Wills:97}, is another variant of the binary polymer model. It is defined as follows.

\subsubsection{The molecule set}
The molecule set $X= X(n)$ consists of all bit strings up to (and including) a maximum length $n$:
\[ X = \{0,1\}^{\leq n} \]
Therefore, there are $|X(n)| = 2^{n+1}-2$ molecule types.

\subsubsection{The food set}
The food set $F$ consists of the two monomers (single bits), i.e., $t=1$:
\[ F = \{0,1\}. \]

\subsubsection{The reaction set}
The reaction set $\R=\R(n)$ consists of additions (ligations) of a monomer to an already existing polymer (bit string) that has a length smaller than the maximum length $n$. Polymers are considered directional (left to right), and the monomer is added to its end. A distinction is made between adding a $0$ to a $0$, a $0$ to a $1$, a $1$ to a $0$, and a $1$ to a $1$. There are thus four ``categories'' of reactions, as follows:
\begin{enumerate}
  \item $\R_1: b0+0 \rightarrow b00,$
  \item $\R_2: b0+1 \rightarrow b01,$
  \item $\R_3: b1+0 \rightarrow b10,$
  \item $\R_4: b1+1 \rightarrow b11,$
\end{enumerate}
where $b$ is any bit string of length at most $n-2$ (including the empty string), i.e., $b \in \{0,1\}^{\leq n-2}$.

So, there are $|\R(n)| = 2^{n+1}-4$ reactions, and each category contains exactly one-quarter ($2^{n-1}-1$) of these reactions. Reactions are again considered to be bi-directional (i.e., for each ligation reaction, there is the equivalent cleavage reaction).

\subsubsection{The catalysis set}
The catalysis set $C$ is made up of combinations of molecules (bit strings) of maximum length $n$ and reaction categories:
\[ C = \{(x, \R_i) | x \in X, |x| = n, i = 1,2,3,4 \}. \]
Notice that if a maximum-length molecule $x$ catalyzes a reaction category $\R_i$, it catalyzes {\it all} reactions in that category.

In the model, these catalysis events are assigned independently and with equal probability $p(n)$ across all possible $(x, \R_i)$ pairs ($2^n \times 4 = 2^{n+2}$ such pairs).

\subsubsection{RAF sets}
RAF sets in the W-H model can contain reactions from any combination of reaction categories. For example, if the (maximum length) molecule $0\cdots0$ catalyses the reaction category $\R_1$, then there exists an RAF set $\R'=\{0+0 \rightarrow 00, 00+0 \rightarrow 000, \ldots, 0\cdots0+0 \rightarrow 0\cdots00\}$. In other words, all reactions involving only polymers of 0s are included in the RAF set (but not all reactions of $\R_1$). Similarly, if the (maximum length) molecule $0\cdots0$ catalyses the reaction category $\R_4$ and (maximum length) molecule $1\cdots1$ catalyses the reaction category $\R_1$, then there exists an RAF set $\R'=\{0+0 \rightarrow 00, 00+0 \rightarrow 000, \ldots, 0\cdots0+0 \rightarrow 0\cdots00, 1+1 \rightarrow 11, 11+1 \rightarrow 111, \ldots, 1\cdots1+1 \rightarrow 1\cdots11\}$. If all four reaction categories are catalysed by at least one molecule, then the entire reaction set $\R$ becomes an RAF set. An RAF set that contains at least some (but not necessarily all) reactions from exactly $j$ different reaction categories ($j=1,2,3,4$) is hereafter referred to as a $j$-category RAF.

\section{Results}

\subsection{The extended binary polymer model}

\subsubsection{Theoretical results}
We start with a theoretical result that generalises the original result of \citet{Mossel:05} to the extended binary polymer model. First,  we require some definitions and a slight modification of lemma 4.3(iii) of that paper.

Let
\[ \lambda_r(n) = p(n) \cdot  \sum_{x \in X(n)} m(x,r,n). \]
Then $\lambda_r(n)$ is the expected number of molecules that catalyse ligation reaction $r$ (for a given $n$). Notice that, from (\ref{probex}) we have: 
\begin{equation}
\label{equ0}
\lambda_r(n)= \sum_{x \in X(n)} Pr[\E(x,r,n)],
\end{equation}
and if we let $\overline{\lambda}(n)$ be the average of these $\lambda_r(n)$ values over all ligation reactions, then we have:
\begin{equation}
\label{equ1}
\overline{\lambda}(n) = \frac{1}{|\R_{+}(n)|} \sum_{r \in \R_{+}(n)} \lambda_r(n) = \frac{1}{|\R_{+}(n)|} \sum_{r \in \R_{+}(n)}  \sum_{x \in X(n)} \Pr[\E(x,r,n)],
\end{equation}
where $\R_{+}(n)$ is the total set of ligation reactions.

Similarly, if we consider the dual quantities that were the focus of  \citet{Mossel:05}, namely the expected number $\mu_n(x)$ of ligation reactions that molecule $x$ catalyses, and $\overline{\mu}(n)$ the average value of these quantities, then we have:
\begin{equation}
\label{equ2}
\overline{\mu}(n) = \frac{1}{|X(n)|} \sum_{x \in X(n)} \mu_n(x)) = \frac{1}{|X(n)|} \sum_{x \in X(n)}  \sum_{r \in \R_{+}(n)} \Pr[\E(x,r,n)],
\end{equation}
where $X(n)$ is the total set of molecule types. 

Comparing Eqns. (\ref{equ1}) and (\ref{equ2}), and noting that $|\R_{+}(n)|/n|X(n)|$ converges exponentially quickly to $1$ with increasing values of $n$ (from Eqn. (2) and (3) in \citet{Mossel:05}, with $\kappa = 2$), we obtain the following asymptotically exact link between these two averages:
$$\overline{\lambda}(n) \approx \overline{\mu}(n)/n.$$

We now state the modified lemma as follows.

\begin{lemma}
\label{helpslem}
Under the extended binary polymer model, the probability that a ligation reaction $r$ is catalysed by at least one molecule is:
$$1-\prod_{x \in X(n)}(1-\Pr[\E(x,r,n)]) \geq 1-\exp(-\lambda_r(n)). $$
\end{lemma}
{\em Proof:} $1-\Pr[\E(x,r,n)]$ is the probability that $r$ is not catalysed by $x$, and so, by the independence assumption,
$\prod_{x \in X(n)}(1-\Pr[\E(x,r,n)])$ is the probability that no molecule catalyses $r$. If we now apply the inequality:
$$\prod_{i} (1-y_i) \leq \exp(-\sum_i y_i),$$
which holds when the $y_i$ values are all non-negative, and invoke Eqn. (\ref{equ0}), the probability that $r$ fails to be catalysed by any molecule is, at most,
$$\exp(-\sum_{x \in X(n)} \Pr[\E(x,r,n)]) = \exp(-\lambda_r(n)).$$
The lemma now follows.

Finally, the generalised result can now be stated as the following theorem. Its proof follows a parallel argument to that provided for proposition 4.4 (ii) of \citet{Mossel:05}, based on Lemma~\ref{helpslem}.

\begin{theorem}
Given an instance $\Q_n$ of the binary polymer model with food set $F$, suppose that for all reactions $r \in \R$, $\lambda_r(n) \geq \lambda$. Then the probability that $\Q_n$ contains an RAF involving all molecules is at least $f(\lambda) = 1-\frac{2(2^{-\lambda})^t}{1-2e^{-\lambda}}$, which is independent of $n$ and which converges to 1 exponentially fast as $\lambda$ increases.
\end{theorem}

This theorem, together with the fact that $\overline{\mu}(n) \approx n\overline{\lambda}(n)$, implies that in the extended binary polymer model, there is also a linear (in $n$)  upper bound on the growth rate in the level of catalysis ($\overline{\mu}(n)$) required to get RAF sets with high probability.   The argument from \citet{Mossel:05} that provides a linear (in $n$) lower bound on the growth rate in the level of catalysis required to get RAF sets also applies here, too.   In other words, adding constraints on which molecules can catalyse which reactions (in the form of $m(x,r,n)$), does not change this main result.

Next, we consider the question of whether the level of catalysis required to get RAF sets with high probability in the extended binary polymer model can be predicted from the (observed) required levels in the original ({\tt RAND}) model. In \citet{Hordijk:12}, we showed that this is possible for the {\tt TMPL} version of the model by using an analytical approximation based on a mathematical technique called the {\it transfer matrix method}. This technique provides a way to calculate  the number of bit strings of a certain length that contain a given substring analytically.  From this, the probability can be derived that an arbitrary bit string (of length $n$ at most, or exactly of length $n$) matches the reaction template of an arbitrary reaction. These probabilities are then used as analytical approximations $\hat{m}(x,r,n)$ of $\overline{m}(x,r,n)$, with which the required probability $p(n$) can be predicted (see \citet{Hordijk:12} for details of this analytical calculation).

Generalising this to the extended binary polymer model, we have:
\[ \overline{\mu}(n) = |\R_{+}(n)| \cdot p(n) \cdot \overline{m}(x,r,n). \]
For the purely random model ({\tt RAND}), $\overline{m}(x,r,n) = 1$, and the required value for $p(n)$ to get, say, a probability $P_n=0.5$ to find RAF sets can be obtained from the simulation results \citep{Hordijk:04}. This provides a corresponding value for the average number of reactions catalysed per molecule, $\overline{\mu}(n)$. Now, for the other model versions, we assume that a similar value for $\overline{\mu}(n)$ is required to get a similar probability $P_n$ of finding RAF sets. However, for these alternative model versions, $\overline{m}(x,r,n) < 1$, and so one would expect that the required value for $p(n)$ needs to increase relative to that in the random model. Using the notation $p(n)$ for the observed required probability in the random model and $\hat{p}(n)$ for the predicted (or expected) required probability in the alternative model ({\tt TMPL}, {\tt MLEN}, or {\tt BOTH}), we then get:
\[ |\R_{+}(n)| \cdot p(n) = |\R_{+}(n)| \cdot \hat{p}(n) \cdot \hat{m}(x,r,n), \]
and thus:
\[ \hat{p}(n) = \frac{p(n)}{\hat{m}(x,r,n)}. \]
Table \ref{tab:mhat} gives the analytically calculated values $\hat{m}(x,r,n)$ (using the calculations described in \citet{Hordijk:12}) for several values of $n$ for the model versions {\tt TMPL}, {\tt MLEN}, and {\tt BOTH}.

\begin{table}[htb]
\begin{center}
\begin{tabular}{rrrr}
\hline
$n$ & {\tt TMPL} & {\tt MLEN}  & {\tt BOTH} \\
\hline
 8 & 0.155 & 0.500 & 0.095 \\
 9 & 0.201 & 0.500 & 0.118 \\
10 & 0.245 & 0.500 & 0.140 \\
11 & 0.288 & 0.500 & 0.161 \\
12 & 0.329 & 0.500 & 0.181 \\
13 & 0.368 & 0.500 & 0.200 \\
\hline
\end{tabular}
\caption{Analytical estimates $\hat{m}(x,r,n)$ (rounded to three digits) for various values of $n$ for the different model versions.}
\label{tab:mhat}
\end{center}
\end{table}

\subsubsection{Computational results}
\begin{figure}[htb] 
\centering
\includegraphics[scale=0.4, angle=270]{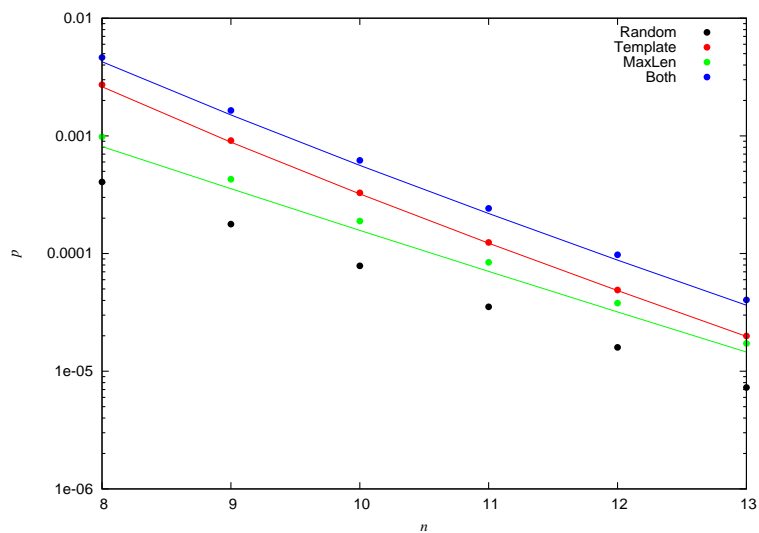}
\caption{The analytically predicted (solid lines) and empirically observed (dots) values for the required probability $p(n)$ (given the respective constraints $m(x,r,n)$) to get $P_n \approx 0.5$ (over 1000 instances) for the various model versions. Note that for the {\tt RAND} model, there are only observed values (dots), on which the analytical predictions for the other model versions are based.}
\label{fig:prob1}
\end{figure}

Figure \ref{fig:prob1} shows the corresponding predicted values $\hat{p}(n)$ for these models (represented by the solid lines), based on the observed values $p(n)$ for the random model ({\tt RAND}; the black dots in Fig. \ref{fig:prob1}). To see how accurate the predicted values $\hat{p}(n)$ are, we performed computer simulations with the alternative models as well;   the observed $p(n)$ values (to get $P_n \approx 0.5$) are shown with dots in the same figure.

\begin{figure}[htb] 
\centering
\includegraphics[scale=0.4, angle=270]{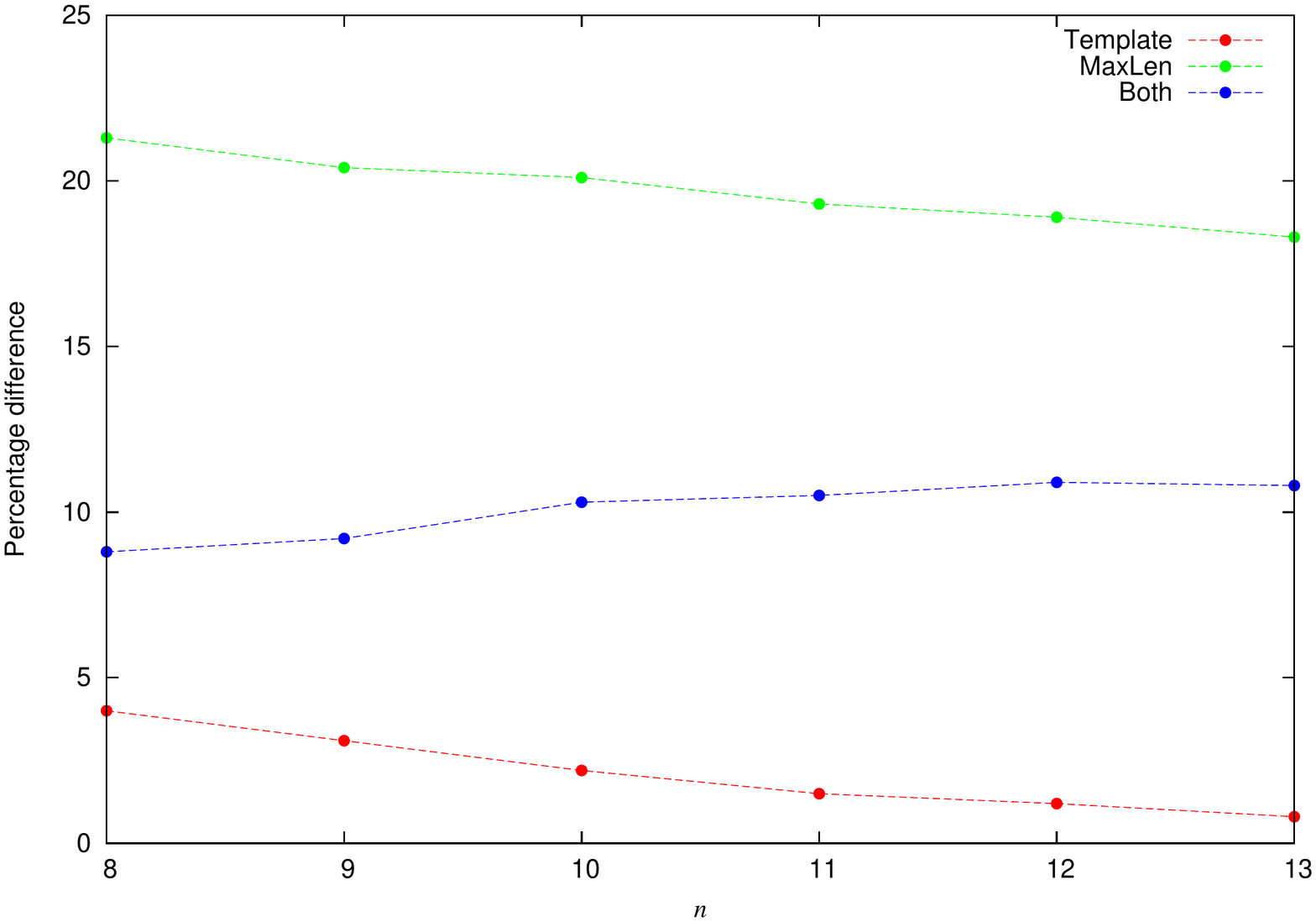}
\caption{The percentage difference between the theoretical and empirical values of $p(n)$ in Fig. \ref{fig:prob1} for the various constrained model versions.}
\label{fig:perc}
\end{figure}

There are clear differences in the prediction accuracy between the different model versions. Figure \ref{fig:perc} shows these differences in terms of the percentage of the predicted values. As the figure shows, the predictions for the {\tt TMPL} (template-based) model are the most accurate, increasingly so for larger values of $n$. This confirms the observation already made in \citet{Hordijk:12} that larger molecules have a higher chance of matching a given (fixed-length) template, somewhere along their sequence. Therefore, for larger values of $n$, the template matching requirement becomes less and less of a constraint, and the predicted values for $p(n)$ get more and more accurate.

The {\tt MLEN} model is the least accurate, but also improves somewhat for larger values of $n$. It is, however, not surprising that for this model, the predictions are less accurate. Using the analytical approximation $\hat{m}(x,r,n)$ implies that this probability is independent and identical for each $(x,r)$ pair. Obviously, this assumption is violated to a large extent in this model version, where only the largest molecules with a length of exactly $n$ can be catalysts (which comprise exactly half of all molecules; hence $\hat{m}(x,r,n) = 0.5$ for all $n$, as shown in Table \ref{tab:mhat}).

Finally, the combination of template-based and maximimum-length catalysis (the {\tt BOTH} model) is somewhere in between in terms of accuracy. Interestingly, the accuracy actually {\it decreases} with larger $n$, but seems to level off eventually. This can be explained by the fact that, over all strings that match a given reaction template, the fraction of maximum-length strings is larger than 0.5 for smaller values of $n$, but converges to 0.5 with increasing $n$. For example, for $n=8$, this fraction is 0.612 but for $n=13$, it decreases to 0.534. The maximum-length requirement becomes, therefore, more of a constraint for larger values of $n$.

This last observation suggests an interesting measure for how much of a structural constraint a given requirement (such as template-based or maximum-length catalysis) imposes on the system in terms of its ability to form RAF sets. The (percentage) discrepancy between the analytically predicted value and the corresponding empirically observed value of $p(n)$ can be taken as a measure of the severity of the imposed constraint. The more a given $m(x,r,n$) distribution deviates from being uniform over all $(x,r)$ pairs, the larger the imposed structural constraint to form RAF sets will be, and, supposedly, the larger the discrepancy between the predicted and observed $p(n)$ values. We return to this issue below by considering the ``constructability'' of RAF sets.

\subsection{The Wills--Henderson (W-H) model}

\subsubsection{Theoretical results}
We start again with some theoretical results, in particular on the probability of RAF sets existing in the W-H model. First, some definitions are required.

Given a subset $\R'$ of $\R$, let $J(\R') = \{j \in \{1,2,3,4\}: \R' \cap \R_j \neq \emptyset\}$ denote the categories of reactions that are represented by at least one reaction in $\R'$.  For a subset $J$ of $\{1,2,3,4\}$, let $P^J_n$ be the probability that the W-H model (for polymers of  length up to  $n$) has an RAF $\R'$ with $J(\R')=J$.

Recall that a {\it $j$-category} RAF is defined as an RAF $\R'$ that contains at least some (but not necessarily all) reactions from exactly $j$ different reaction categories ($j=1,2,3,4$).  
Thus, the probability that the W-H model (for polymers of length up to $n$)   has a $j$-category RAF is:
 $$P_n^{(j)} =  \sum_{J \subseteq \{1,2,3,4\}: |J|= j}P^J_n,$$
This probability the model has an RAF  is then $\sum_{j}^4P_n^{(j)}$.
We now derive theoretical approximations for these various probabilities.  

\begin{itemize}

\item {\bf 1-category RAFs}\\
If $J=\{2\}$ or $J=\{3\}$, only strings of length two can be created from the food set; therefore $P_n^J=0$ in both these cases.

If $J = \{1\}$ or $J=\{4\}$ then we generate exactly one sequence $x$ of length $n$ (either the all-0 string or the all-1 string). In this case, we have:
$$P^J_ n = \PP( x \mbox{ catalyses } \R_j)\PP(x \mbox{ doesn't catalyse any $\R_k$ for $k \neq j$})$$
$$= p(n)(1-p(n))^3,$$
and so
$$P_n^{(1)} \sim \binom{4}{1} p(n) = 4p(n).$$

\item {\bf  2-category RAFs}
If $|J|=2$, say $|J| =\{i,j\}$, then in all cases, exactly two sequences, $x$ and $x'$, of length $n$ can be generated from the food set. Thus, in this case, $P^J_ n$ equals:
$$\PP( \mbox{ $x$ or $x'$ catalyses } \R_i)\PP( \mbox{ $x$ or $x'$ catalyses }  \R_j) \PP(\mbox{$x$ and $x'$  doesn't catalyse $\R_k$ or $\R_l$} ),$$
where $\{k,l\} = \{1,2,3,4\} - \{i,j\}$.  Therefore,
$$P^J_n= (1-(1-p(n))^2)(1-(1-p(n))^2)(1-p(n))^4   \sim 4p(n)^2.$$
Thus we have:
$$P^{(2)}_n \sim \binom{4}{2} \cdot 4 p(n)^2 = 24 p(n)^2.$$

\item {\bf  3-category RAFs}
The dominant and most interesting case is where $J=\{1,2,3\}$ or $J=\{2,3,4\}$. Consider the first possibility (the other is similar). Here the number of molecules of maximal length $n$ we can generate is precisely the number of sequences of length $n$ in which two `1's never appear consecutively (i.e., $\cdot  \cdot  \cdot  11 \cdot  \cdot  \cdot $ is forbidden).  It is a classical result in enumerative combinatorics that this number is simply the Fibonacci number $F_{n+1}$ where $F_0=F_1=1$ and $F_{n} = F_{n-1}+F_{n-2}$ for all $n>1$. The easiest way to see this is by virtue of an alternative description of the Fibonacci recursion:
$$F_{n+1}=2F_{n}-F_{n-2}.$$
Thus we have:
$$P^J_n = \left[1-(1-p(n))^{F_{n+1}}\right]^3(1-p(n)) \sim (1-e^{-c\mu})^3,$$
where $p(n) = \mu/((1+\sqrt{5})/2)^n$ and $c= \lim_{n \rightarrow \infty} F_{n+1}/\mu^n$ (notice that $(1+\sqrt{5})/2 = 1.618$, the `golden ratio').

The other  3-category RAFs are for $J=\{1,2,4\}$ and $\{1,3,4\}$. For these the number of molecules of maximal length that are generated from $F$ grows (only) linearly in $n$. So the 3-category RAFs are dominated in probability by the two  interesting cases above.

\item {\bf  4-Category RAFs}
A  4-category RAF exists if and only if the entire set $\R=\cup_{j=1}^4 \R_j$ of all reactions is an RAF.  For the `only if' part of this claim, note that if  $\R'$ is an RAF with $J(\R')=\{1,2,3,4\}$ then all of the reactions in $\R$ are catalysed, and all the  reactants of  $\R$ can be constructed from the food set $F=\{0,1\}$ using $\R$. Thus, for  $J=\{1,2,3,4\}$:
$$P^{J}_ n = \prod_{i=1}^4 \PP(\R_i \mbox{ is catalysed by at least one polymer of length $n$ })$$
$$=  \prod_{i=1}^4 \left [1-\PP(\R_i \mbox{ is not catalysed by any polymer of length $n$ } )\right]$$
$$=(1-(1-p(n))^N)^4,$$
for $N=2^n$. Let us now write $p(n) = \lambda /N$. Then we have:
$$ P^{(4)} = (1-e^{-\lambda})^4 +o(1),$$
where $o(1)$ refers to a term that converges exponentially quickly  to 0 as $n$ increases. Thus, for $P^{(4)}_n = 0.5$, we have $\lambda = -\ln(1-2^{-0.25}) = 1.838..$, and therefore:
$$p(n) \sim 1.838/2^n,$$
where (here and below) $\sim$ denotes asymptotic equivalence as $n$ grows.
\end{itemize}

Notice that when  $p(n) = 1.838/2^n$ (the 0.5 threshold for a  4-category RAF), we have the following:
\begin{itemize}
\item the probability of a 1-category RAF is $\sim 7.35/2^n$;
\item the probability of a 2-category RAF is $\sim 81/4^n$ (a much smaller probability than for a 1-category RAF);
\item the probability of a 3-category RAF of the most probable type (i.e. $J=\{1,2,3\}$ or $J=\{2,3,4\}$) is 
$P^J_n \sim (1.838\frac{F_{n+1}}{2^n})^3$ which converges to 0 exponentially quickly with $n$ (but much slower than for a 1-category or 2-category RAF).
\end{itemize}

In summary, the $P_n=0.5$ threshold for an RAF in the W-H model converges asymptotically (and exponentially quickly with $n$) to the $P_n=0.5$ threshold for 4-category RAFs. Any RAFs that are not 4-category RAFs are most likely to be 3-category RAFs. Of the remaining two, a 1-category RAF is much more probable than a  2-category RAF (but still much less than  3-category RAF). Thus the ordering is:

\bigskip

4-category $>>$ 3-category $>>$ 1-category $>>$ 2-category.

\bigskip

Finally, we consider the existence of irreducible RAF sets. In \citet{Steel:13}, we showed that, in general, finding the {\it smallest} irrRAFs is a hard problem, and we introduced a randomised algorithm to find (arbitrary) irrRAFs and sample their sizes. However, in the specific case of the W-H model, it is actually possible to construct a polynomial-time algorithm to find the size of the smallest possible irrRAFs (they need not be unique) within a maxRAF $\R'$:
\begin{enumerate}
\item Take the set $M$ of molecules of maximum length $n$ that catalyse at least one reaction in $\R'$.
\item For each {\it minimal} subset $S$ of $M$ that includes exactly one catalyst for each catalysed reaction category (so $S$ is a subset of four molecules at most from $M$), do the following:
  \begin{enumerate}
  \item For each $x \in S$, let $R(x)$ be the sequence of $n-1$ reactions that generates $x$ from $F$ by adding monomers;
  \item Take $R_S = \bigcup_{x \in S} R(x)$.
  \end{enumerate}
\item The size of the smallest possible irrRAFs of $\R'$, is the size of the smallest set $R_S$ generated in Step 2.
\end{enumerate}
Note that this algorithm is polynomial in $|M|$ (the number of catalysts in $\R'$). Moreover, the algorithm implies that the size of the smallest irrRAFs of $\R'$ must lie between $n-1$ and $4(n-1)$.

\subsubsection{Computational results}

\begin{figure}[htb] 
\centering
\includegraphics[scale=0.4, angle=270]{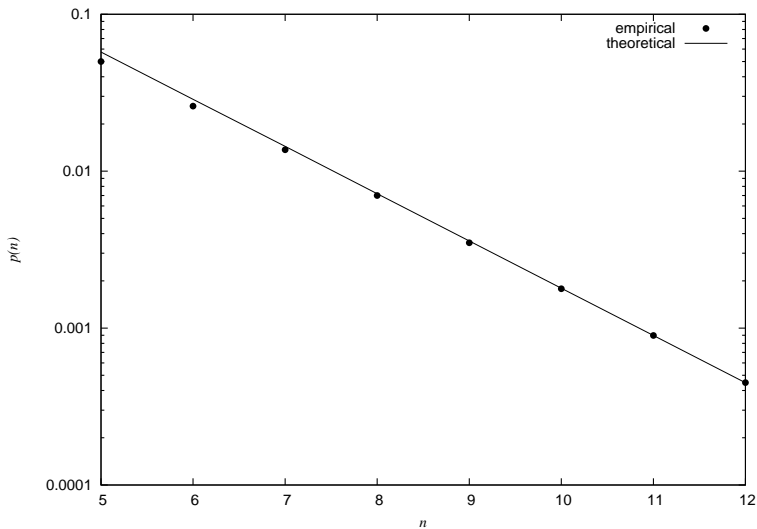}
\caption{The theoretically predicted (solid line) and empirically observed (dots) values for the required probability $p(n)$ to get $P_n \approx 0.5$ (over 1000 instances) in the W-H model.}
\label{fig:prob}
\end{figure}

We performed computer simulations with the W-H model for various values of $n$ to check the accuracy of the theoretical predictions. Figure \ref{fig:prob} shows the results, where the solid line represents the theoretical values and the dots the empirically observed values for $p(n)$ to get a probability of around $P_n=0.5$ to find RAF sets (averaged over 1000 instances). As the plot shows, the theoretical predictions are very accurate, and increasingly so for larger values of $n$.

Furthermore, to check the prediction on the ordering of the four categories in terms of their likelihood, Table \ref{tab:perc} shows the percentage of RAF sets for $n=8$ (and a value of $p(n)$ that gives $P_n=0.5$) that are $j$-category RAFs for $j=1,2,3,4$. Indeed, 4-category RAFs dominate, the remainder consisting of 3-category, 1-category and 2-category RAFs (in that order), as predicted. However, for $n > 10$, basically all RAFs that are found are 4-category RAFs.

\begin{table}[htb]
\centering
\begin{tabular}{|l|cccc|}
\hline
$j$ & 1 & 2 & 3 & 4 \\
\hline
\% & 0.786 & 0.196 & 1.768 & 97.250 \\
\hline
\end{tabular}
\caption{The percentage of instances where the found RAF set contains $j=1, 2, 3, 4$ reaction categories in the W-H model with $n=8$ and $p(n)=0.0070$.}
\label{tab:perc}
\end{table}

\begin{figure}[htb] 
\centering
\includegraphics[scale=0.5]{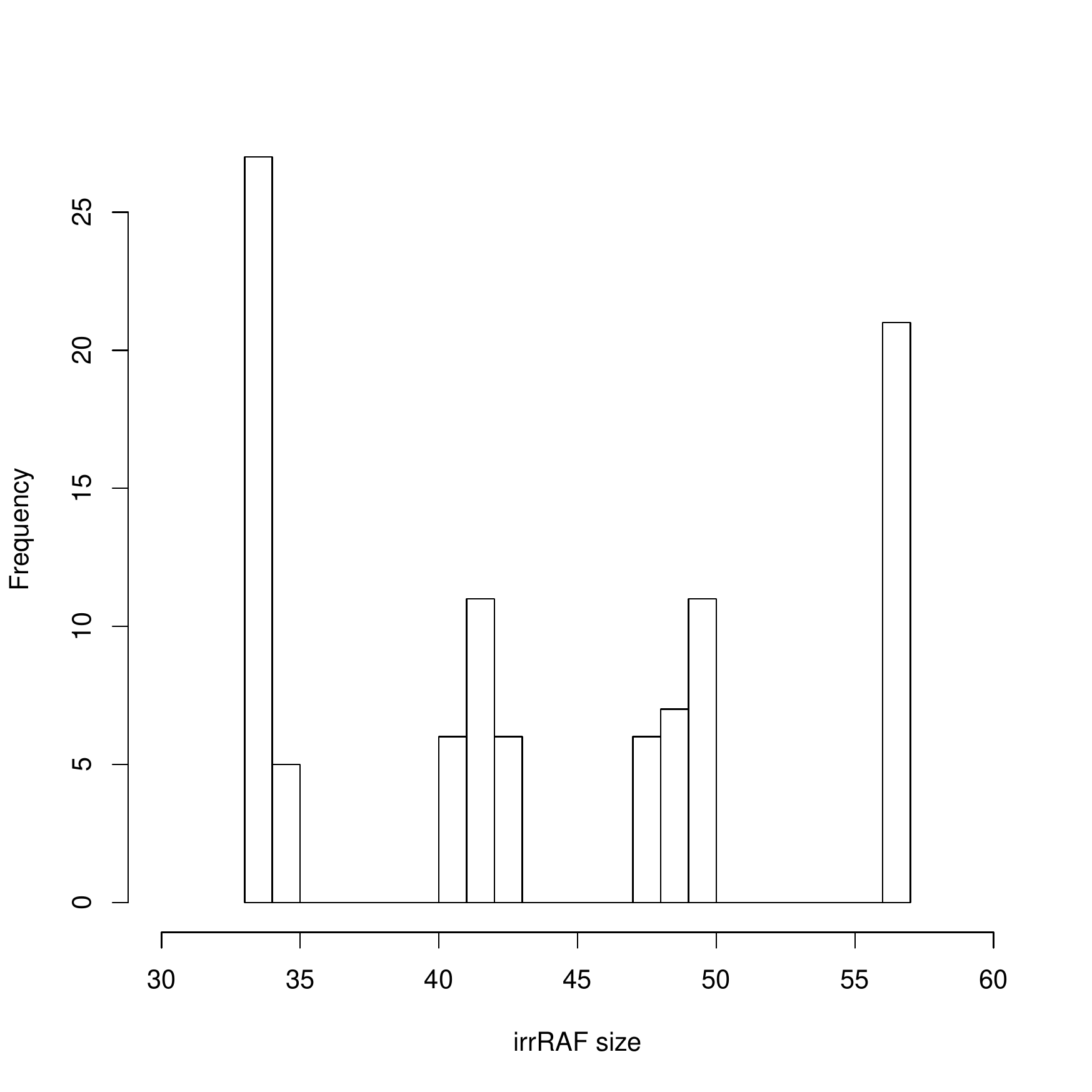}
\caption{Histogram of the sizes of 100 randomly generated irrRAF sets within one particular 4-category maxRAF in the W-H model with $n=10$.}
\label{fig:hist}
\end{figure}

Even though the maximal RAF sets in the W-H model are predominantly 4-category RAFs, consisting of the entire reaction set $\R = \R_1 \cup \R_2 \cup \R_3 \cup \R_4$, they do contain many smaller RAF subsets. Figure \ref{fig:hist} shows a histogram of the sizes (in number of reactions) of 100 irrRAFs as found by the randomised algorithm \citep{Steel:13} within one particular maximal RAF set for $n=10$. The size of this 4-category maxRAF is $|\R'| = |\R(10)| = 2^{11}-4 = 2044$ reactions. However, as the figure shows, the sizes of the  irrRAFs found range from 33 to 57 reactions (i.e., they are much smaller than the maxRAF).  Indeed, the smallest irrRAF size found by the randomised algorithm (33 reactions) is equal to the minimum irrRAF size calculated by the exact algorithm for the W-H model introduced above. 

\subsection{Constructability of RAFs}

Note that the formation of an RAF starting from the food set (of polymers up to length $t$)  requires a minimum of $\log_2(n)-t$ reactions to proceed uncatalysed before the first catalyst can be produced in the {\tt MLEN} version of the binary polymer model. Similarly, there need to be at least $n-1$ such uncatalysed reactions in the W-H model. The definition of RAF sets allows for this to occur, since reactions can still proceed uncatalysed, albeit at a much slower rate. However, a chemical network in which catalysts are produced before they are needed is likely to have a significant advantage over the types where catalysis comes late (as in the {\tt MLEN} model), because in the former case, an RAF would form more quickly, and before the reactants dissipate. In the extreme case, we have the notion of a `constructively autocatalytic F-generated set' (CAF), studied in \citet{Mossel:05}, which can be built up in such a way that each reaction is catalysed by molecules already available. A more formal definition follows.

Given a chemical reaction system, $\Q= (X, \R, C)$ with food set $F$, recall that $\R'$ is a CAF if there is a linear ordering of $\R'$,  $r_1,r_2, \ldots$, so that, for each $i>1$:
\begin{itemize}
\item[{\bf (P1)}] all reactants of $r_i$ are contained in the closure of $F$ relative to $\{r_1,\ldots, r_{i-1}\}$;
\item[{\bf (P2)}] at least one catalyst of $r_i$ lies in the closure of $F$ reative to $\{r_1,\ldots, r_{i-1}\}$. 
\end{itemize}

If $\R'$ is an RAF but not a CAF, an interesting question then is whether or not there exists an ordering that satisfies (P1), and which requires at most $k$ violations of (P2). When such an ordering exists, we say that the RAF is {\em constructible from $F$ modulo $k$ catalysations}.

{\bf Example:}\\
The maxRAF $\{r_1,r_2,r_3\}$ shown in Fig. \ref{fig:CRS} is constructible from $F$ modulo one catalysation. For example, the ordering $r_2, r_1, r_3$ fails (P2) for just the first reaction ($r_2$), and clearly satisfies (P1).

Now consider the following decision problem.

\bigskip

\indent{{\bf $k$-cat-RAF}\\
INSTANCE:  A chemical reaction system and food set $(\Q, F)$,
an RAF $\R'$ for $(\Q, F)$ and a positive integer $k$.\\
QUESTION: Is $\R'$ constructible from $F$ modulo $k$ catalysations?}

\bigskip

Determining whether a given RAF is constructible from $F$ modulo $k$ catalysations turns out to be an intractable problem, as the following result shows. The proof is provided in the Appendix.

\begin{theorem}
\label{hardthm}
$k$-cat-RAF is NP-hard.
\end{theorem}

The importance of this theorem is that it tells us that, rather than searching for a general exact algorithmm for min-k-RAF, we should consider special cases, or try to obtain upper and lower bounds for the solution that can be calculated efficiently. For example, an easily computatable upper bound on the smallest value of $k$ for which $\R'$ is constructible from $F$ modulo $k$ catalaysations is to construct a nested sequence $$F=X_0 \subset X_1 \subset X_2 \cdots \subset X_m = F \cup \pi(\R')$$ of subsets of $X$ in which $X_{i+1}$ (for $0\leq i < m$) is the set of molecules in $X_m$ that can be generated from reactants in $X_i$ by applying a reaction from $\R'$. Let us say that a molecule $x \in X_{i+1}$ is {\em premature} if $x$ is not in $X_i$ and if none of the reactions from $\R'$  that generate $x$ from reactants in $X_i$ is catalysed by any molecule in $X_{i}$. Then an upper bound on the smallest value of $k$ for which $\R'$ is constructible from $F$ modulo $k$ catalaysations is the sum of the number of premature molecules in the sequence $X_1 \subset X_2 \cdots  \subset X_m$.

\section{Concluding comments}

In this paper, we have studied the consequences of constraining the formation of self-sustaining autocatalytic (RAF) sets of polymers so that the variation of the catalysts' properties with their structure conforms with the main features of chemistry that underpin the maintenance of functional specificity in molecular biological systems. The original binary polymer model, as first described by \citet{Kauffman:71, Kauffman:86} and later refined by others, provides important insights into the probability of RAF formation. However, the simplicity of the binary polymer model comes at the price of biochemical realism. It gives a very short molecule, or one with no matching template or other generically defined features conferring recognition capability, the same probability of catalysing a given cleavage-ligation reaction as a long sequence. An exactly matching  template or ``keyhole" active site is the sort of structure most likely capable of precisely discriminating substrates.  Thus, it is important to ask how  results derived from the simple binary polymer model might be affected if the action of catalysis were more specifically dictated by  the fit between the reaction and the potential catalyst.

Here, we have investigated two types of extensions of the binary polymer model. 

The first is the extended binary polymer model, for which a catalyst is required to have either a matching template or to be of maximal length, or subject to both these requirements. The maximum length model (MLEN) represents an extreme case of models which are constrained by the plausible demand that longer molecules have higher probability of catalysing a reaction than shorter ones; we study this extreme case, as we expect to find the greatest difference from the original binary polymer model. In the extended binary polymer models, we demonstrated that the degree of catalyzation required for the likely emergence of RAFs grows linearly with the length of the sequences, as has  already been established for the original binary polymer model \citep{Mossel:05}.

We then asked whether or not we can predict the density of catalytic funtionality in the polymer sequence space required for the emergence of RAFs in these extended models, by substituting in its place the simple binary polymer model with the density adjusted to match the degree of catalysation of the corresponding more complex model averaged over all reactions. Calculating these average densities of catalytic function is possible (by standard methods from combinatorics) and, for the  template-matching model, the predicted degree of catalyzation required for RAF formation can be estimated quite precisely (with a discrepancy of approximately 1\% at $n=13$) by the surrogate binary polymer model. For the other two models, the discrepancy is higher (20\% for the MLEN model and 10\% for both combined).  

The reason for this increased discrepancy may be explained, at least in part, by the increased heterogeneity of the distribution of catalysts in the polymer sequence space for these latter two models. For example, when only maximum-length molecules are catalysts (the MLEN model), we can exactly fit the expected degree of catalyzation in this model using a simple binary polymer model (with the degree of catalyzation chosen appropriately) but the distribution of catalysis in the MLEN model shows higher variance than that in the surrogate simple model. More precisely, suppose we select a molecule $\chi$ uniformly at random and, conditional on $\chi=x$, consider the number $N(x)$ of reactions that molecule $x$ catalyses. Consider the variance of the compound random variable $N(\chi)$.  Under the MLEN model, this variance $\sigma^2_{{\rm MLEN}}$, is greater than the variance  $\sigma^2_{{\rm RAND}}$ of a matching simple binary polymer (where the average degree of catalysation per molecule in both models is $m$). This can be seen by comparing the following equations\footnote{The first equation holds because in RAND, and for every molecule $x$, $N(x)$ has a Poisson distribution with mean -- and therefore variance -- equal to $m$; the second equation is from  the identity $Var[N(\chi)] = E[Var[N|\chi]] + Var[E[N|\chi]]$, together with the fact that half the molecules have maximal length.}:

$$\sigma^2_{{\rm RAND}} = m    \mbox{ and } \sigma^2_{{\rm MLEN}} = m +m^2.$$

The second extension we investigated was the  W-H model in which the reactions are more restrictive than the original binary polymer model.  Instead of allowing molecules to combine freely by ligation operations, the reactions in the W-H model attach just one monomer at a time to a polymer; moreover,  catalysis is possible only by maximal length molecules, with each such molecule having the same probability of catalysing any one of the four classes of ligation reactions (i.e. $\cdots x  + y \rightarrow \cdots xy$ for $x,y= 0,1$).   

The W-H model has an advantage over the other models in that one can mathematically calculate the exact probability that an RAF set exists, and specify its $j$-category type. Also, one can compute the size of the smallest RAF exactly, which was recently shown to be an NP-hard problem for the simple binary polymer model \citep{Steel:13}. Moreover, the smallest RAF in the W-H model is always small (linear in $n$) but for the random binary polymer model, it was recently proved that, at the level of catalysis where RAFs are starting to emerge, the smallest RAFs are almost certain to be of a size that is exponential in $n$ \citep{Steel:13}.

These attractive features of the W-H model are tempered by the rather coarse way in which RAFs emerge -- these typically include all of  the reactions (or initially at least one of the four classes of reactions). Thus, the probability of achieving system specificity of the type common in biological systems, i.e., catalysis of a selection of reactions, rather than all them, is very small. The original consideration of the W-H model \citep{Wills:97} focused on the bias away from a random distribution of catalytic functions in the sequence space of maximum length polymers (the ratio $f'/f$ in the nomenclature of that paper) needed for an  $i$-category RAF to survive dynamic competition with $k$-category RAFs for $k>i$.

Although W-H systems with an unbiased (random) distribution of catalyzations in the MLEN sequence space have a strong tendency toward the maximal RAF set, much smaller RAF subsets will generally exist. The specification of these small RAFs in terms of reaction sets is rather artificial, in that any molecules catalysing a reaction specific to a small RAF will also catalyse all of the other reactions in the same category. However, the existence of small RAFs demonstrates the small number of exactly specified reactions, perhaps coincidental byproducts of some other catalytic process of the same broad genre, that are needed to seed the generation of the maximal RAF. It is a feature of both approaches (the extended binary polymer model and the W-H model) that a number of reactions must proceed uncatalysed, or as a result of other processes external to the system in which autocatalysis eventually occurs, until the catalysts that contribute to the RAF are formed. This is most obvious for the W-H model, where a minimum of $n-1$ such uncatalysed reactions are required.    

For the extended binary polymer model in which only maximal length molecules are catalysts we need a minimum of $\log_2(n)-t$ uncatalysed reactions to generate the molecules capable of maintaining an RAF. Such steps represent obstacles to the formation of an RAF; reactions that are uncatalysed can proceed, but only at a slow rate, and this may be too slow in the presence of dissipation or degrading side-reactions. Thus it would be helpful to be able to compute, for any RAF, the smallest number of reactions that need to proceed uncatalysed before the RAF can be established. Our final result was to show that this problem is NP-hard, so it is unlikely that a polynomial-time algorithm exists for it, and suggests that alternative strategies should instead be explored. As a first example, we described a simple upper bound on the the number of uncatalysed reactions required, which  can be computed in polynomial time.

It should be possible to extend the results further; some extensions would likely be straightforward and  lead to similar results (for example, polymers over a non-binary alphabet, where results are typically similar to the binary case \citep{Mossel:05}), while other extensions would probably introduce new complications (for example, allowing molecules to inhibit reactions, or introducing degrading side-reactions). The dynamics of RAF sets, which have been studied for the simple binary polymer model \citep{Hordijk:12b}, would also be of interest in these extended models. These and other studies should help provide increasing biological relevance of RAF theory, with the ultimate aim of providing a better understanding of how the first primitive self-sustaining autocatalytic systems may have become established.

\section*{Acknowledgments}

We thank the {\em Allan Wilson Centre for Molecular Ecology and Evolution} and the {\em Alexander von Humboldt Foundation} for helping fund part of this research.

\bibliographystyle{spbasic}
\bibliography{tuebingen_model}

\begin{thebibliography}{22}
\providecommand{\natexlab}[1]{#1}
\providecommand{\url}[1]{{#1}}
\providecommand{\urlprefix}{URL }
\expandafter\ifx\csname urlstyle\endcsname\relax
  \providecommand{\doi}[1]{DOI~\discretionary{}{}{}#1}\else
  \providecommand{\doi}{DOI~\discretionary{}{}{}\begingroup
  \urlstyle{rm}\Url}\fi
\providecommand{\eprint}[2][]{\url{#2}}

\bibitem[{Crick(1958)}]{Crick:58}
Crick FHC (1958) On protein synthesis. Symposia of the Society for Experimental
  Biology 12:138--163

\bibitem[{Crick(1970)}]{Crick:70}
Crick FHC (1970) Central dogma of molecular biology. Nature 227:561--563

\bibitem[{Eigen(1971)}]{Eigen:71}
Eigen M (1971) Self-organization of matter and the evolution of biological
  macromolecules. Naturwissenschaften 58:465--523

\bibitem[{Eigen and Schuster(1979)}]{Eigen:79}
Eigen M, Schuster P (1979) The Hypercycle. Springer, Berlin

\bibitem[{Garey and Johnson(1979)}]{Garey:79}
Garey MR, Johnson DS (1979) Computers and Intractability: A Guide to the Theory
  of NP-Completeness. W. H. Freeman

\bibitem[{Hordijk and Steel(2004)}]{Hordijk:04}
Hordijk W, Steel M (2004) Detecting autocatalytic, self-sustaining sets in
  chemical reaction systems. Journal of Theoretical Biology 227(4):451--461

\bibitem[{Hordijk and Steel(2012{\natexlab{a}})}]{Hordijk:12b}
Hordijk W, Steel M (2012{\natexlab{a}}) Autocatalytic sets extended: Dynamics,
  inhibition, and a generalization. Journal of Systems Chemistry 3:5

\bibitem[{Hordijk and Steel(2012{\natexlab{b}})}]{Hordijk:12}
Hordijk W, Steel M (2012{\natexlab{b}}) Predicting template-based catalysis
  rates in a simple catalytic reaction model. Journal of Theoretical Biology
  295:132--138

\bibitem[{Hordijk and Steel(2013)}]{Hordijk:13}
Hordijk W, Steel M (2013) A formal model of autocatalytic sets emerging in an
  {RNA} replicator system. Journal of Systems Chemistry 4:3

\bibitem[{Hordijk et~al(2011)Hordijk, Kauffman, and Steel}]{Hordijk:11}
Hordijk W, Kauffman SA, Steel M (2011) Required levels of catalysis for
  emergence of autocatalytic sets in models of chemical reaction systems.
  International Journal of Molecular Sciences 12(5):3085--3101

\bibitem[{Hordijk et~al(2012)Hordijk, Steel, and Kauffman}]{Hordijk:12a}
Hordijk W, Steel M, Kauffman S (2012) The structure of autocatalytic sets:
  Evolvability, enablement, and emergence. Acta Biotheoretica 60(4):379--392

\bibitem[{Kauffman(1971)}]{Kauffman:71}
Kauffman SA (1971) Cellular homeostasis, epigenesis and replication in randomly
  aggregated macromolecular systems. Journal of Cybernetics 1(1):71--96

\bibitem[{Kauffman(1986)}]{Kauffman:86}
Kauffman SA (1986) Autocatalytic sets of proteins. Journal of Theoretical
  Biology 119:1--24

\bibitem[{Kauffman(1993)}]{Kauffman:93}
Kauffman SA (1993) The Origins of Order. Oxford University Press

\bibitem[{Mossel and Steel(2005)}]{Mossel:05}
Mossel E, Steel M (2005) Random biochemical networks: {T}he probability of
  self-sustaining autocatalysis. Journal of Theoretical Biology 233(3):327--336

\bibitem[{Schr{\"o}dinger(1944)}]{Schroedinger:44}
Schr{\"o}dinger E (1944) What is Life? Cambridge University Press, Cambridge

\bibitem[{Steel(2000)}]{Steel:00}
Steel M (2000) The emergence of a self-catalysing structure in abstract
  origin-of-life models. Applied Mathematics Letters 3:91--95

\bibitem[{Steel et~al(2013)Steel, Hordijk, and Smith}]{Steel:13}
Steel M, Hordijk W, Smith J (2013) Minimal autocatalytic networks. Journal of
  Theoretical Biology 332:96--107

\bibitem[{Vaidya et~al(2012)Vaidya, Manapat, Chen, Xulvi-Brunet, Hayden, and
  Lehman}]{Vaidya:12}
Vaidya N, Manapat ML, Chen IA, Xulvi-Brunet R, Hayden EJ, Lehman N (2012)
  Spontaneous network formation among cooperative {RNA} replicators. Nature
  491:72--77

\bibitem[{Vasas et~al(2012)Vasas, Fernando, Santos, Kauffman, and
  Sathm{\'a}ry}]{Vasas:12}
Vasas V, Fernando C, Santos M, Kauffman S, Sathm{\'a}ry E (2012) Evolution
  before genes. Biology Direct 7:1

\bibitem[{Watson and Crick(1953)}]{Watson:53}
Watson JD, Crick FHC (1953) Genetical implications of the structure of
  deoxyribonucleic acid. Nature 171:964--967

\bibitem[{Wills and Henderson(2000)}]{Wills:97}
Wills P, Henderson L (2000) Self-organisation and information-carrying capacity
  of collectively autocatalytic sets of polymers: ligation systems. In: Bar-Yam
  Y (ed) Unifying Themes in Complex Systems: Proceedings of the First
  International Conference on Complex Systems, Perseus Books, pp 613--623

\end{thebibliography}

\section*{Appendix: Proof of Theorem~\ref{hardthm}}

{\em Proof:} We will reduce the graph theoretic problem VERTEX COVER to $k$-cat-RAF (a similar reduction was employed in \citet{Steel:13} for a quite different problem).   Recall that for a graph $G=(V,E)$, a {\em vertex cover} of $G$ is a subset $V'$ of $V$ with the property that each edge of $G$ is incident with at least one vertex in $V'$; VERTEX COVER has as its instance a graph $G=(V,E)$ and an integer $K$ and we ask whether or not $G$ has a vertex cover of size at most $K$. This is a well-known NP-complete problem \citep{Garey:79} (indeed, it is one of Karp's original 21 NP-complete problems). Given an instance $(G=(V,E), K)$ of VERTEX COVER we show how to construct an instance $(X_G, \R_G, C_G, F_G, k)$, of $k$-cat-RAF for which the answers to the two decision problems are identical (here, $\R_G$ is an RAF).

First we construct $F_G$ and $X_G$. For each $v\in V$ let $a_v, b_v$ be two distinct elements of $F_G$ and let $x_v$ be an element of $X_G-F_G$. Order $E$ as $e^1, \ldots, e^{|E|}$ and for each $j=1, \ldots, |E|$ let $d_j$ be a distinct element of $F$ and $y_j$ an element of $X_G-F_G$. Let $d_0$ be another distinct element of $F_G$.   Thus $F_G$ consists of the $2|V|+|E|+1$ elements:
$$F_G:= \{d_j: 0 \leq j \leq |E|\} \cup \{a_v, b_v: v \in V\},$$
$X_G-F_G$ consists of $|V|+ |E|$ elements:
$$X_G-F_G:= \{x_v: v \in V\} \cup \{ y_j: 1\leq j \leq |E|\}.$$

\begin{figure}[htb] 
\centering
\includegraphics[scale=0.7]{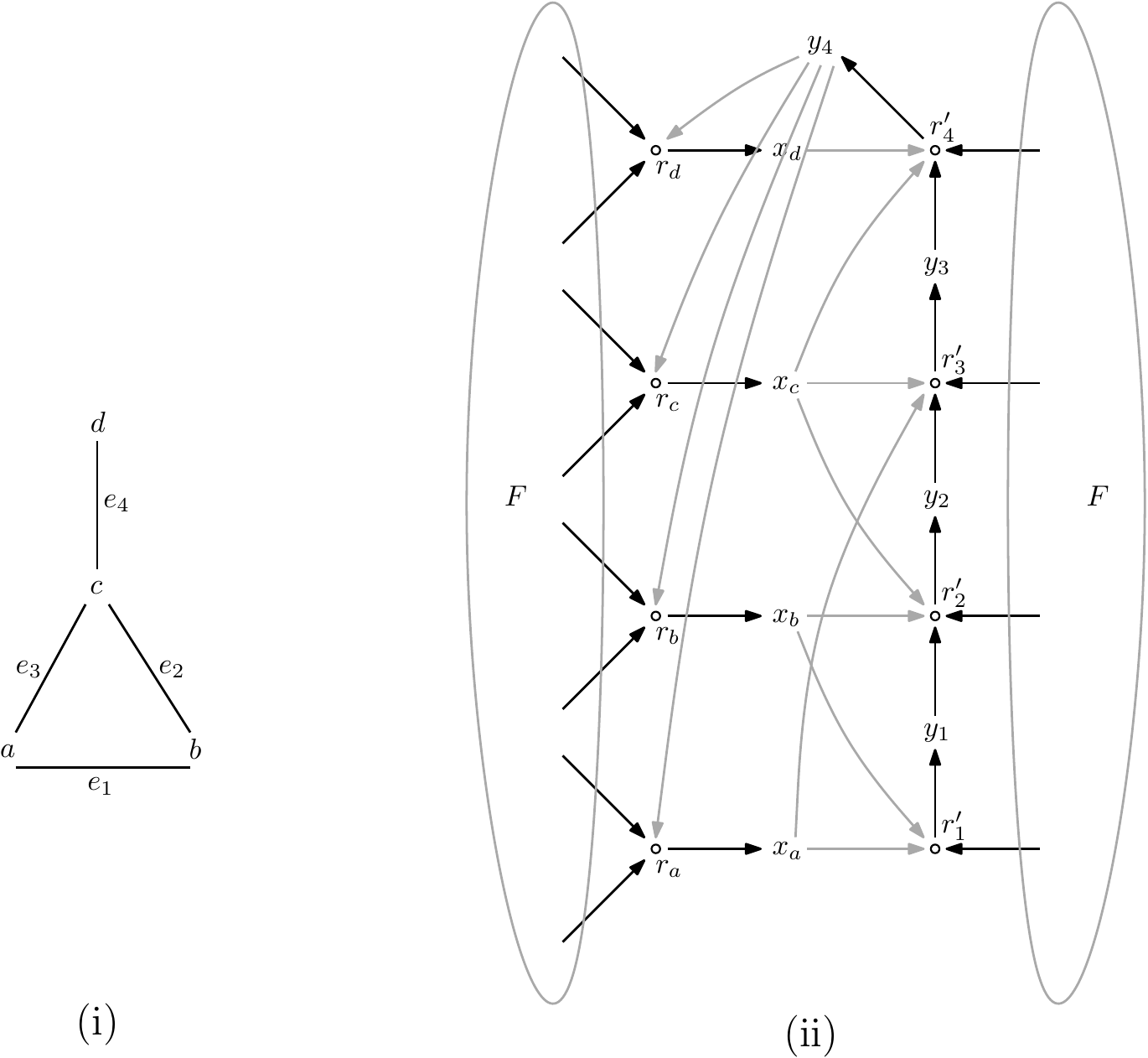}
\caption{(i) A graph $G$ and (ii) the associated CRS $\Q_G$, consisting of 8 reactions that form a RAF, and with the super-catalyst ($y_4$) at the top. }
\label{figure0}
\end{figure}

For each $v \in V$, define a reaction $$r_v: a_v + b_v \rightarrow x_v.$$
For  each $1<j \leq |E|$,  define the reaction:
$$r'_j:  y_{j-1}+d_j \rightarrow y_j,$$
and for  $j=1$ let:
$$r'_1: d_0+d_1 \rightarrow y_1.$$ 
For any subset $U$ of $V$,  let 
$\R_U = \{r_v: v \in U\}$, let  $$\R_V := \{r_v: v \in V\} \mbox{ and } \R_E := \{r'_j: 1\leq j \leq |E|\},$$
and set $\R_G = \R_V \cup \R_E.$
Thus we have specified  $X_G, F_G$ and $\R_G$ and it remains to define the catalysis ($C_G$) assignment, which is as follows:
\begin{itemize}
\item
If $e^j = (u^j, v^j)$ (where $u^j, v^j \in V$)  then $r'_j$ is catalysed by both $x_{u^j}$ and $x_{v^j}$ (but by no other molecules).  
\item
In addition, each reaction $r_v: v\in V$ is catalysed by $y_{|E|}$ and by no other molecule -- we call the molecule  $y_{|E|}$ the {\em super-catalyst}.
\end{itemize}

An example of this construction is illustrated in Fig.~\ref{figure0}. We have now fully specified the catalysation and thereby the pair $(\Q_G, F_G)$ constructed from $G$ ($\Q_G = (X_G, \R_G, C_G)$).

{\bf CLAIM:} A finite graph $G$ has a vertex cover of size at most $K$ if and only there is an ordering of $\R_G$ that satisfies (P1) and  involves at most $K$ violations of (P2).

To establish this claim, first suppose that $V'$ is a vertex cover of $G$ of size at most $K$. Then order $\R_{V'}$ arbitrarily and place these as the first reactions in a linear ordering, followed by the reactions in $\R_E$ in the order $r'_1, \ldots, r'_{E}$, and finally the remaining reactions in $\R_{V-V'}$ in arbitrary order as the final segment of the ordering. This ordering just requires $|V'|=K$ violations of (P2) for the initial reactions (i.e. $\R_{V'}$), and it also satisfies (P1), and so provides the required ordering of $\R_G$.

Conversely, suppose that there is an ordering of $\R_G$, $r_1, \ldots, r_{|V|+|E|}$ that satisfies (P1) and involves at most $K$ violations of (P2). Let $J$ denote the set of $j$ for which (P2) fails for $r_j$, and let 
$$J_V = \{j\in J: r_j \in \R_V\} \mbox{ and } J_E = \{j\in J: r_j \in \R_E\}.$$
Each  $j \in J_E$ corresponds to some edge $e$ of $G$,  so we will let $v(j)$ denote any vertex of $G$ incident with $e$.

Now, $\{r_j: j \in J_V\} \cup \{r_{v(j)}: j \in J_E\}$ is a subset of $\R_V$ of size at most $K$, and so corresponds to $\R_{V'}$ for a subset $V'$ of $V$ of size at most $K$.  We show that $V'$ is an edge cover of $G$, by showing that any given edge $e$ contains at least one vertex from $V'$.

First, observe that the reaction $r'_e \in \R_E$ is one of the reactions $r_j$ in the above ordering of $\R$. We consider two cases: (i) $j \in J_E$ and 
(ii) $j \not\in J_E$. In Case (i), $v(j) \in V'$ and so $e$ contains this vertex from $V'$. In Case (ii), $r_j$ is catalysed by a product of a reaction $r_i$ that appears earlier in the ordering. This implies that $r_i = r_v$ for some vertex $v$ of $V$; therefore,  if $v \in V'$ then $e$ contains an element of $V'$. It remains to consider the case where $v \not\in V'$ (i.e.,  $i \not\in J_V$). We  will show that this case never arises  by deriving a contradiction on the assumption that it does. If $i \not\in J_V$ then $r_i$ is catalysed by a reaction $r_k$ that appears earlier than $i$ in the given ordering of $\R$. However, the only  reaction that can catalyse $r_i$ is $r'_{|{E}|}$, which  must therefore appear as $r_k$ for some $k<i$ in the ordering (since we are assuming that $i \not\in J_V$). Summarising, we have:
\begin{equation}
\label{chain}
k < i < j.
\end{equation}
It is at this point that we invoke (P1). Notice that the reactants for $r_k$ do not become available until all the other reactions in $\R_E$ -- including $r_j$ -- have occurred. By (P1), this requires that $j<k.$ Combining this with Inequality~(\ref{chain}) we obtain the required contradiction required to exclude the last case. This completes the proof.

\end{document}